\documentclass[a4paper,11pt]{article}

\usepackage{jinstpub} 

\usepackage{graphicx}
\usepackage{subfigure}
\usepackage{multirow}
\usepackage{lineno}

\title{\boldmath Design of Guide Tube Calibration System for JUNO Experiment}

\author[a]{Yuhang Guo,}
\author[a,1]{Qingmin Zhang,\note{Corresponding author.}}
\author[b]{Feiyang Zhang,}
\author[c,d]{Mengjiao Xiao,}
\author[b]{Jianglai Liu,}
\author[a]{Eryuan Qu}

\affiliation[a]{School of Nuclear Science and Technology, Xi'an Jiaotong University, Xi'an 710049, China}
\affiliation[b]{School of Physics and Astronomy, Shanghai Jiao Tong University, Shanghai 200240, China}
\affiliation[c]{Department of Physics, University of Maryland, College Park, Maryland 20742, USA}
\affiliation[d]{Center of High Energy Physics, Peking University, Beijing 100871, China}

\emailAdd{zhangqingmin@mail.xjtu.edu.cn}

\abstract{Jiangmen Underground Neutrino Observatory (JUNO) is designed to determine the neutrino mass hierarchy using a 20 kton liquid scintillator detector. To calibrate detector boundary effect, the Guide Tube Calibration System (GTCS) has been designed to deploy a radioactive source along a given longitude on the outer surface of the detector.  In this paper, we studied the physics case of this system via simulation, which leads to a mechanical design.}


\begin{document}
\maketitle
\flushbottom
\def\degree{${}^{\circ}$}

\section{Introduction}
\label{sec:intro}

The Jiangmen Underground Neutrino Observatory (JUNO) is a reactor anti-neutrino experiment currently under construction in Kaiping, China, 53 km away from both the Taishan and Yangjiang Nuclear Power Plants in Guangdong Province. The primary goal of JUNO is to determine the mass ordering of the neutrinos. JUNO's Central Detector (CD) is a Liquid Scintillator (LS) detector enclosed in an acrylic sphere with an inner diameter of 35.4 m and a wall thickness of 12 cm. Ultrapure water will be filled outside of the acrylic sphere, in which region there will be 17000 20-inch and 25000 3-inch photomultipliers (PMTs) viewing the scintillation photons produced by the inversed beta decay events occurred inside CD when electron anti-neutrinos interact with the protons in the LS. In order to determine the neutrino mass hierarchy to more than 3$\sigma$ median sensitivity within 6 years of operation, JUNO detector is required to have very high energy resolution of $3\%/\sqrt{E}$ and an absolute energy scale uncertainty of less than 1\%~\cite{a,b,c,h,n}.

Ideally, a large LS detector is a total absorption calorimeter, with
its total number of detected photoelectrons proportional to the
deposited energy. However, due to geometrical effects and complication
in the photon propagation (e.g. internal reflections at the
acrylic-water surface), the total photoelectron yield is significantly
position dependent, as indicated in Fig.~\ref{fig:The Response
  Distribution in r_direction} (a), particularly for the region close to the detector
boundary. This position non-uniformity, which would otherwise produce
an energy bias and an energy-independent constant term to the energy
resolution, clearly needs to be calibrated and corrected~\cite{b}. To
achieve this, the JUNO calibration system consists of four independent
systems, each of which is capable of deploying radioactive sources into certain
region in the CD, as illustrated in Fig.\ref{fig:The Response
  Distribution in r_direction} (b). In this paper, we will focus on
the design of the Guide Tube Calibration System (GTCS), which brings
radioactive source along a chosen longitude along the CD
boundary. Similar calibration systems are used in experiments such as
Double Chooz~\cite{l} and CUORE~\cite{i}, but the detector size of
JUNO clearly poses additional design challenges.

\begin{figure*}[htp]
	\centering
    \subfigure[]{    
          \label{fig_a}     
          \includegraphics[width=0.8\textwidth]{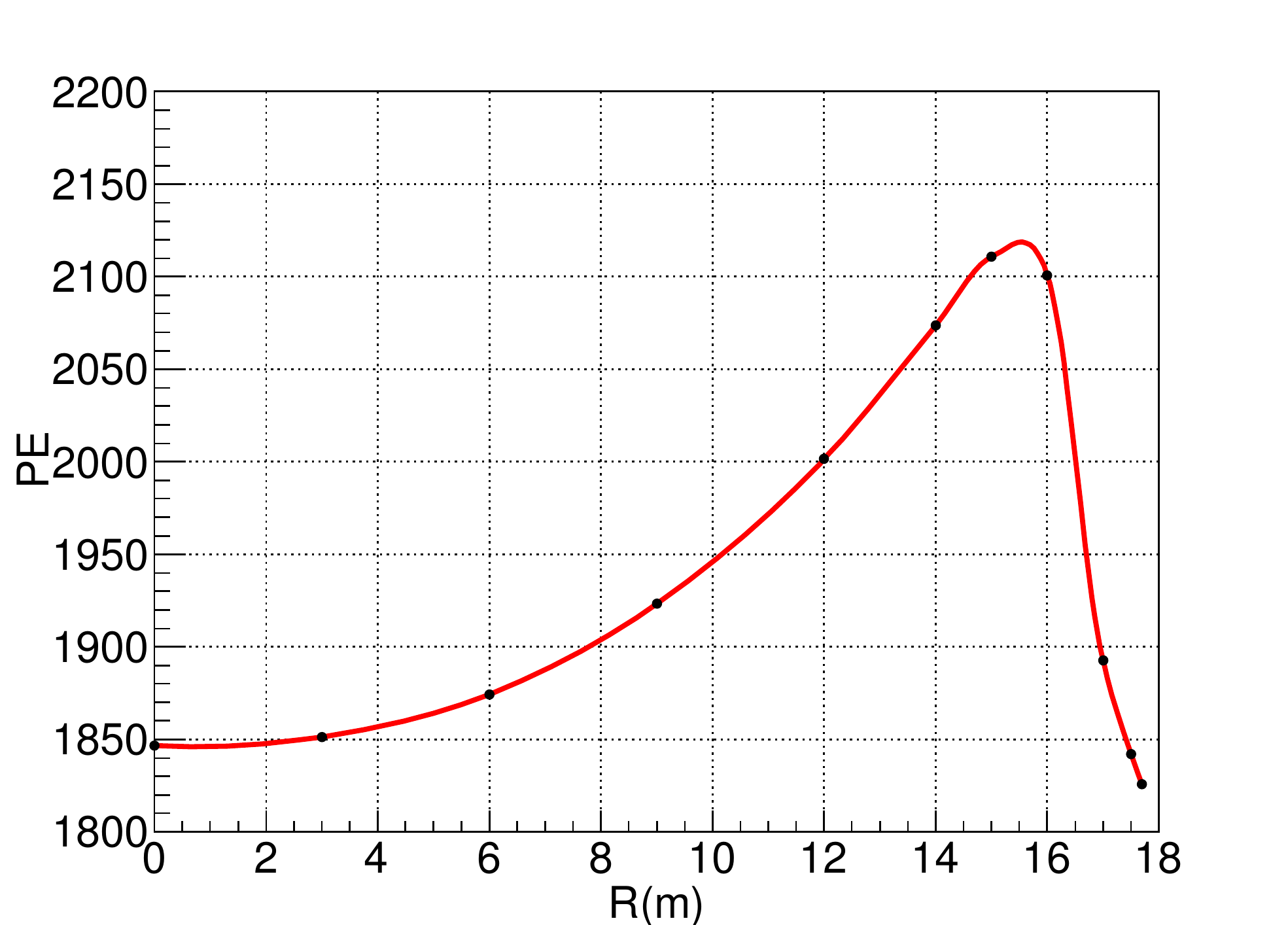}} 
    \subfigure[]{    
          \label{fig_b}     
          \includegraphics[width=0.8\textwidth]{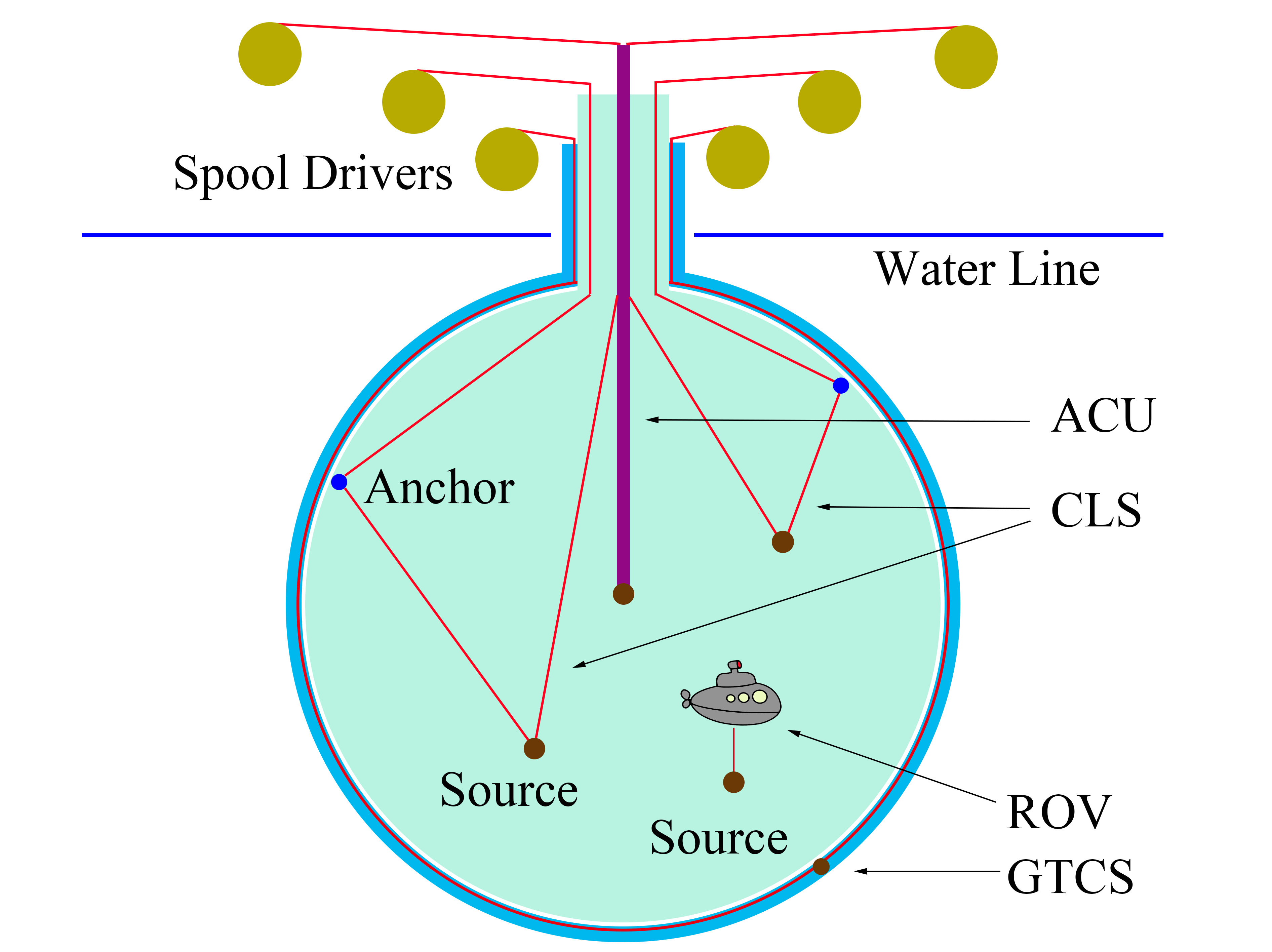}} 
	\caption{(a) Non-uniformity of the photoelectron yield of the
          JUNO CD as a function of radius, simulated with 1.5 MeV electrons. (b)
          Diagram of JUNO Calibration System including the Automatic
          Calibration Unit (ACU), the Cable Loop System (CLS), the
          Guide Tube Calibration System (GTCS) and the
          Remotely-Operated Vehicle (ROV).}
	\label{fig:The Response Distribution in r_direction}
\end{figure*}

The rest of this paper is organized as follows. In the next section, the boundary
response of JUNO will be thoroughly studied by simulation, based on
which we settle on the location of GTCS and the calibration
strategy. Then in Section ~\ref{sec:mechanical}, we discuss the design of the
system as well as some tests with a prototype setup.

\section{Physics Considerations of the GTCS Design}
As mentioned in the introduction, JUNO requires an overall absolute
energy scale uncertainty of 1\%. We therefore require that the systematic
uncertainty of the energy scale calibration at the detector boundary
by GTCS shall reach the same level of 1\%. In this section, we will discuss
simulation studies connecting this basic requirements to the
mechanical design of GTCS. For simplicity, in this paper we discuss
the study using a $^{40}$K gamma source ($\sim$1.46 MeV single gamma
ray) for which the location of the full absorption peak is used to
calibrate the energy scale, but our conclusion remains unchanged if
other gamma or neutron sources are considered.

\subsection{Outside or Inside CD?}
\label{sec:in_or_out}
To mock up a real physics events happening at the detector boundary,
ideally one would like to deploy a gamma source along the inner
surface of the acrylic sphere. However, for mechanical reasons and
cleanliness considerations, deploying the source along a tube on the
outer surface of acrylic is much more feasible. For a nominal gamma
source, the latter option will inevitably lead to a stronger Compton
leakage tail. Therefore the first important issue to study is the
systematic difference in the full absorption peaks in these two options.

The simulation was based on the JUNO offline simulation software
SNIPER with version J17v1r1-pre1, a Geant4-based simulation
framework\cite{j,d}.  To accurately simulate the detector response at
the CD boundary, realistic engineering geometry has been incorporated
into the SNIPER.  The JUNO CD is immersed in water. The acrylic sphere
is mechanically held by the Stainless-steel (SS) latticed Shell with
591 SS connection bars. These bars are distributed along 23 different
latitude lines \cite{f} , and connected to the acrylic nodes on the
acrylic sphere. A SS fastener (SF) is embedded inside each acrylic
node to hold the bar. On top of the CD, a 800~mm penetration mates to
a SS chimney connecting to the calibration house and the LS filling
system. Optical properties of these ``dead'' components are modeled
based on data from bench measurements. We assume that the guide tube is a reflective PTFE tube with a reflectivity of 90\%, an inner diameter of 16 mm and a thickness of 1.5 mm.

\begin{figure*}[htp]
	\centering
    \subfigure[]{    
          \label{fig_a}     
          \includegraphics[width=0.8\textwidth]{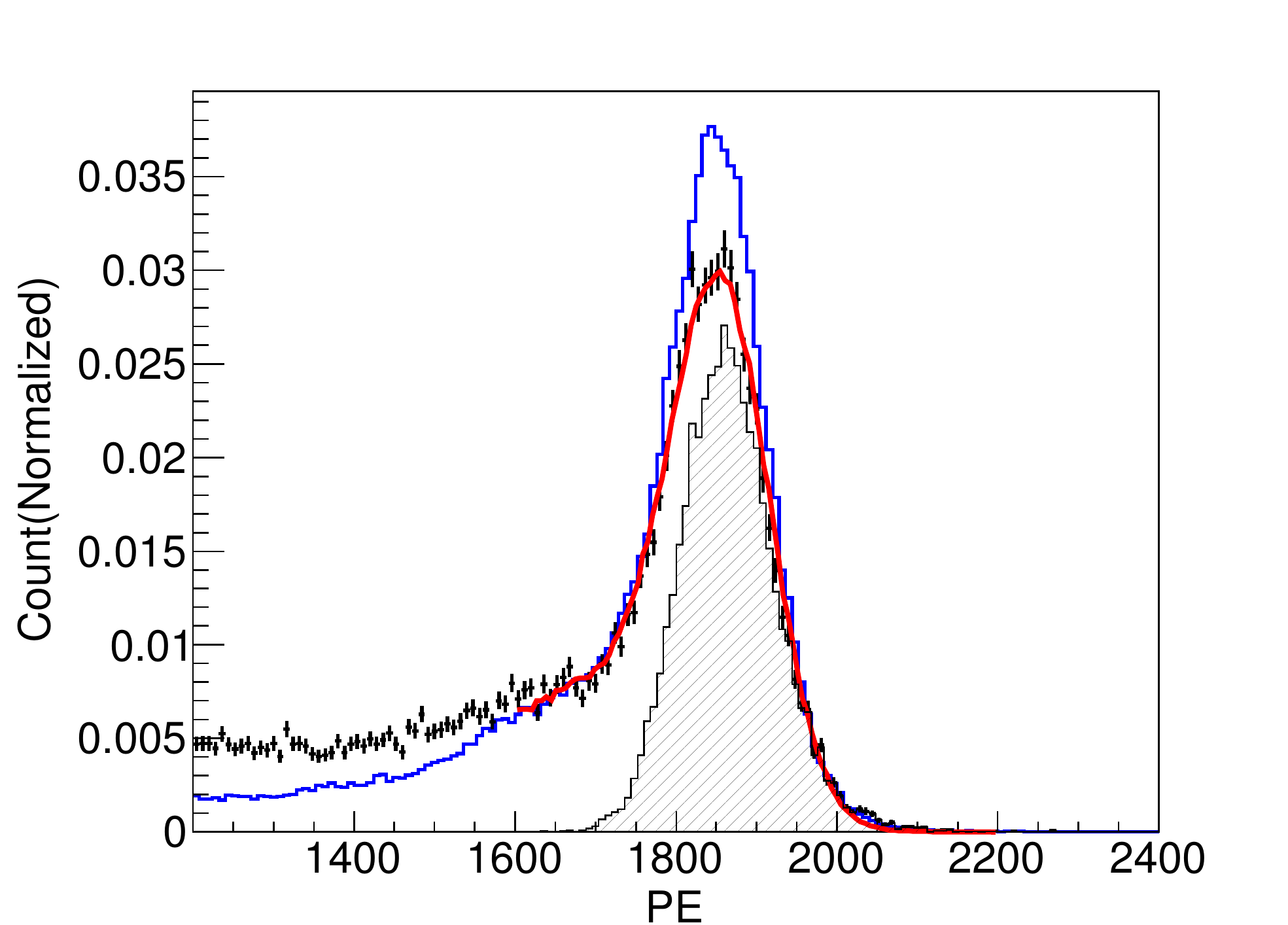}} 
    \subfigure[]{    
          \label{fig_b}     
          \includegraphics[width=0.8\textwidth]{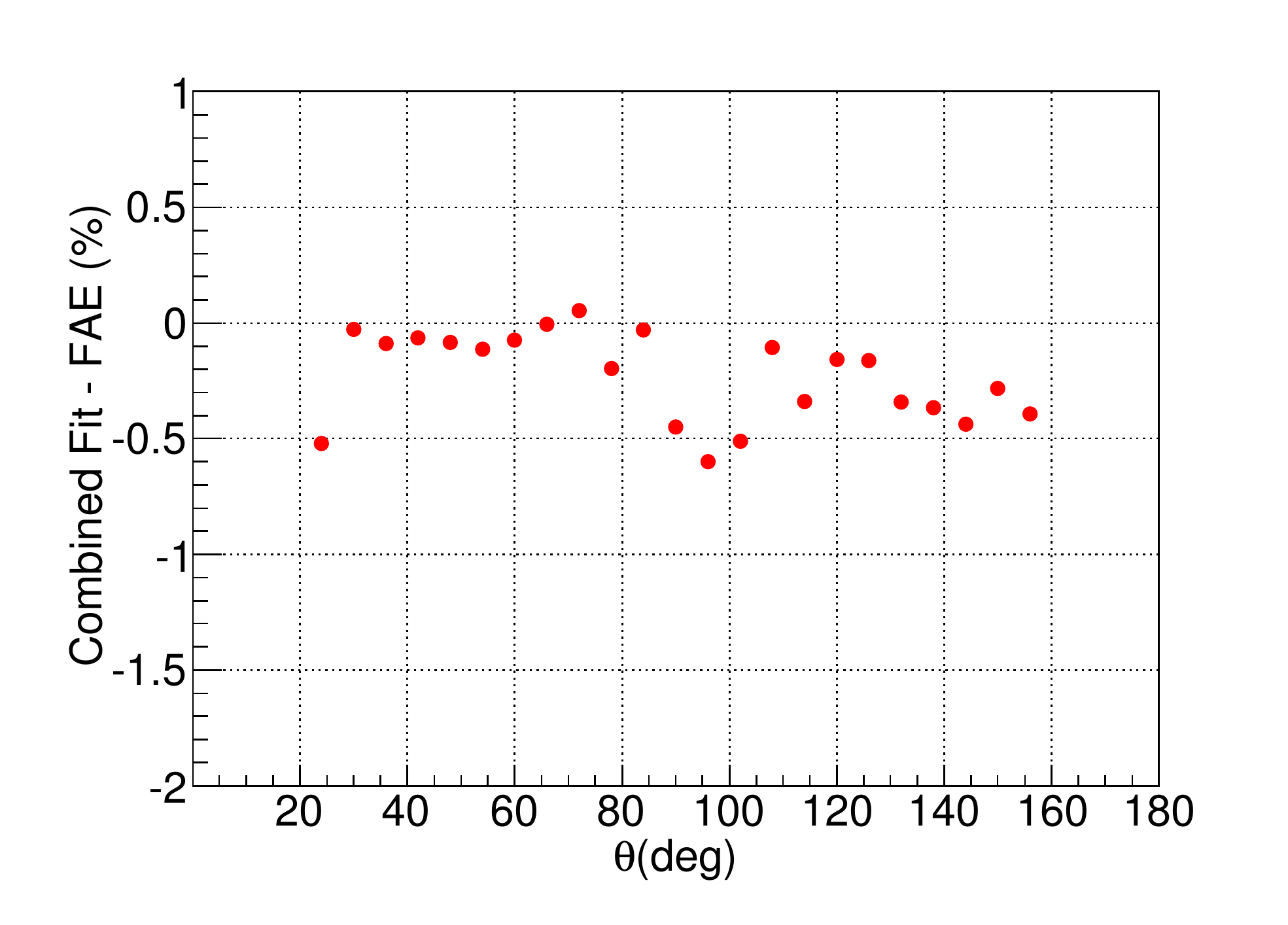}} 
	\caption{(a) Black: The typical PE spectrum from a $^{40}K$ source received by PMTs. Shaded: The corresponded FAE peak. Red: The fitting result with Eq\eqref{eq:fited}. Blue: The PE spectrum corresponding to a naked $^{40}$K source located at CD inner surface. (b) The fractional difference between the combined fitted Gaussian peak and the true FAE peak, indicating the uncertainty of the combined fit. }
	\label{fig:The_PE_Spectrums}
\end{figure*}

The typical total photoelectron (PE) spectrum from a $^{40}$K source
located in a guide tube attached to the outer CD surface is shown in
Fig.~\ref{fig:The_PE_Spectrums} (black histogram). The full absorption
peak is clearly observed, but overlaps with the Compton leakage
tail. For comparison, the spectrum corresponding to a naked $^{40}$K
source located at the CD inner surface is overlaid in the same
figure (blue histogram). In this case the contribution from the tail is reduced by a
factor of two or so, but still significant, as expected. The CD is only
sensitive to the converted gamma energy in the active LS region,
therefore whether the gamma originates from the inner or outer surface
does not make a great difference as in either case there is significant
inactive volume surrounding the gamma vertex.

We fit the observed spectrum by using a combination of a full absorption peak and Compton tail as shown in Eq\eqref{eq:fited},
\begin{equation}
\label{eq:fited}
f_{\rm m}(\rm{PE})=g(\mu,\sigma)+f_{c}(\rm{PE})\,,
\end{equation}
where $g(\mu,\sigma)$ is a Gaussian function with mean $\mu$ and
standard deviation $\sigma$, and $f_{\rm c}(\rm{PE})$ is an average
Compton tail (over an entire latitude line) obtained from simulation,
with the vertical scale floating in the fit. It was verified that for
all locations such a model can produce satisfactory fit. For a nominal 30000 calibration events, the statistical
uncertainty of the centroid of the Gaussian reaches
0.1\%. In the simulation, we can also select fully
absorbed events (FAE) with all gamma energy deposited in the active
region. To evaluate the systematic uncertainty from the combined fit,
the centroid in Fig.~\ref{fig:The_PE_Spectrums} (a) is compared to that
for the FAEs for different locations on the latitude (Fig.~\ref{fig:The_PE_Spectrums} (b)). 
Due to the CD top chimney and bottom flange affecting tube deployment, which will be mentioned in section 3.2 in details, only the points in the range from 24\degree to 156\degree are studied. 
On average, one observes a scattered difference with a range of $\pm$0.6\%, which
serves as an estimate of systematic uncertainty of such fit.

\begin{figure*}[htp]
    \centering
    \includegraphics[width=0.8\textwidth]{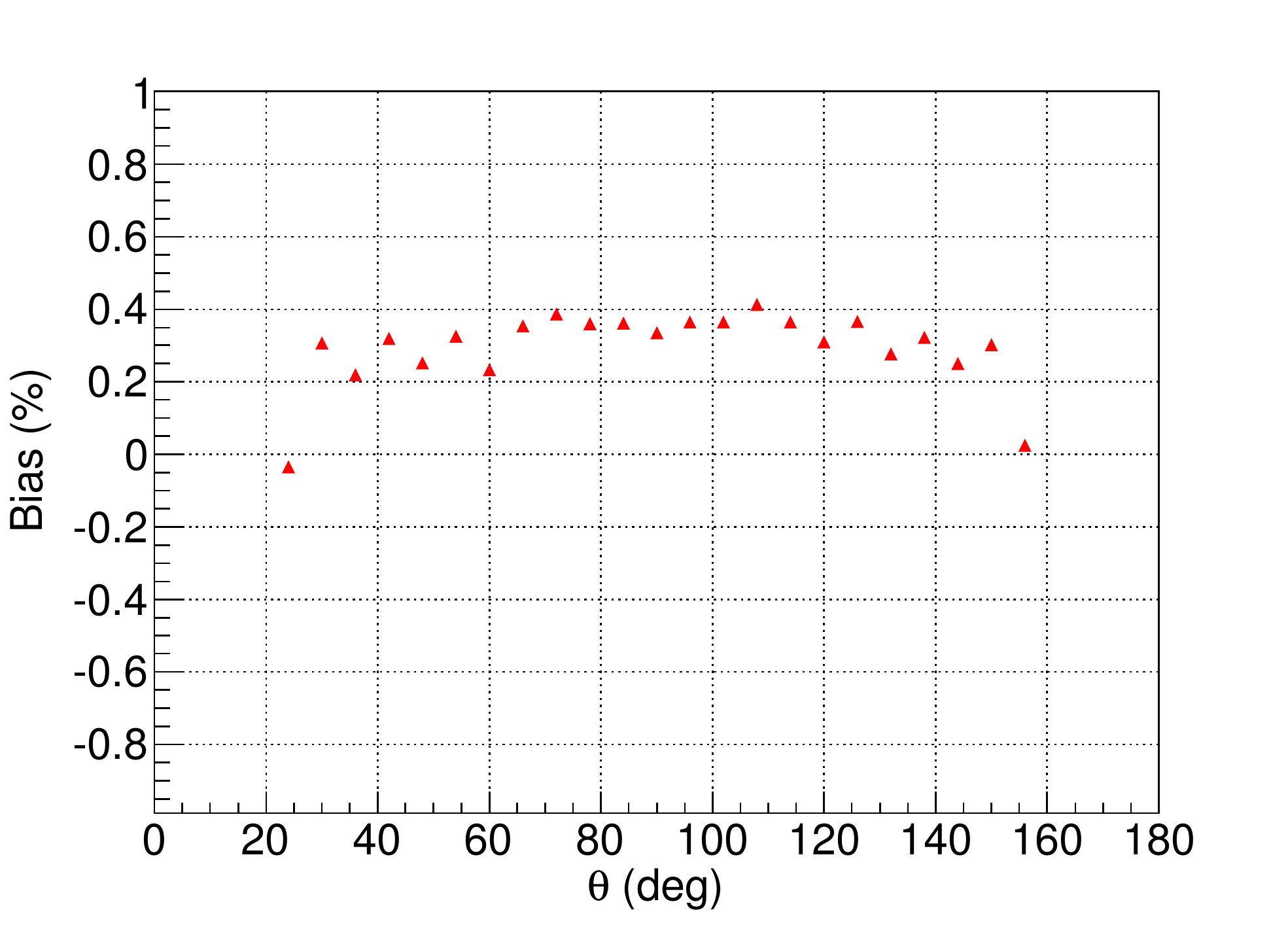}
    \caption{The relative difference between the FAEs when the source in
      is located at the inner (ideal) or outer (practical) surface of
      the CD.}
	\label{fig:LBE}
\end{figure*}

Beside the fit systematics, the mean values of the FAE for the source
located at the inner or outer CD surface is also compared in
Fig.~\ref{fig:LBE} as a function of the azimuthal angle. One sees that
on average the inner response (ideal) would be about 0.3\% higher than that
for the guide tube attached on the outer surface. Such a bias is, in
principle, correctable based on simulation. However, for conservativeness, we take it also
as a source of systematic uncertainty. Therefore, the overall
systematic uncertainty of the FAE peak when the source is located at
the outer surface of the CD is $\sqrt{0.6^2+0.3^2} \approx 0.7\%$.

To conclude from the studies above, the systematic uncertainty to the
FAE for the outer GTCS strategy is estimated to be 0.7\%. This is
satisfactory in comparison to the sub-1\% requirement for the total
energy scale uncertainty. We therefore settle on this option as the baseline design of GTCS.

\subsection{Choosing the Path of the Guide Tube}
\label{sec:average_function}
\begin{figure*}[htp]
	\centering
    \subfigure[]{    
          \label{fig_a}     
          \includegraphics[width=0.80\textwidth]{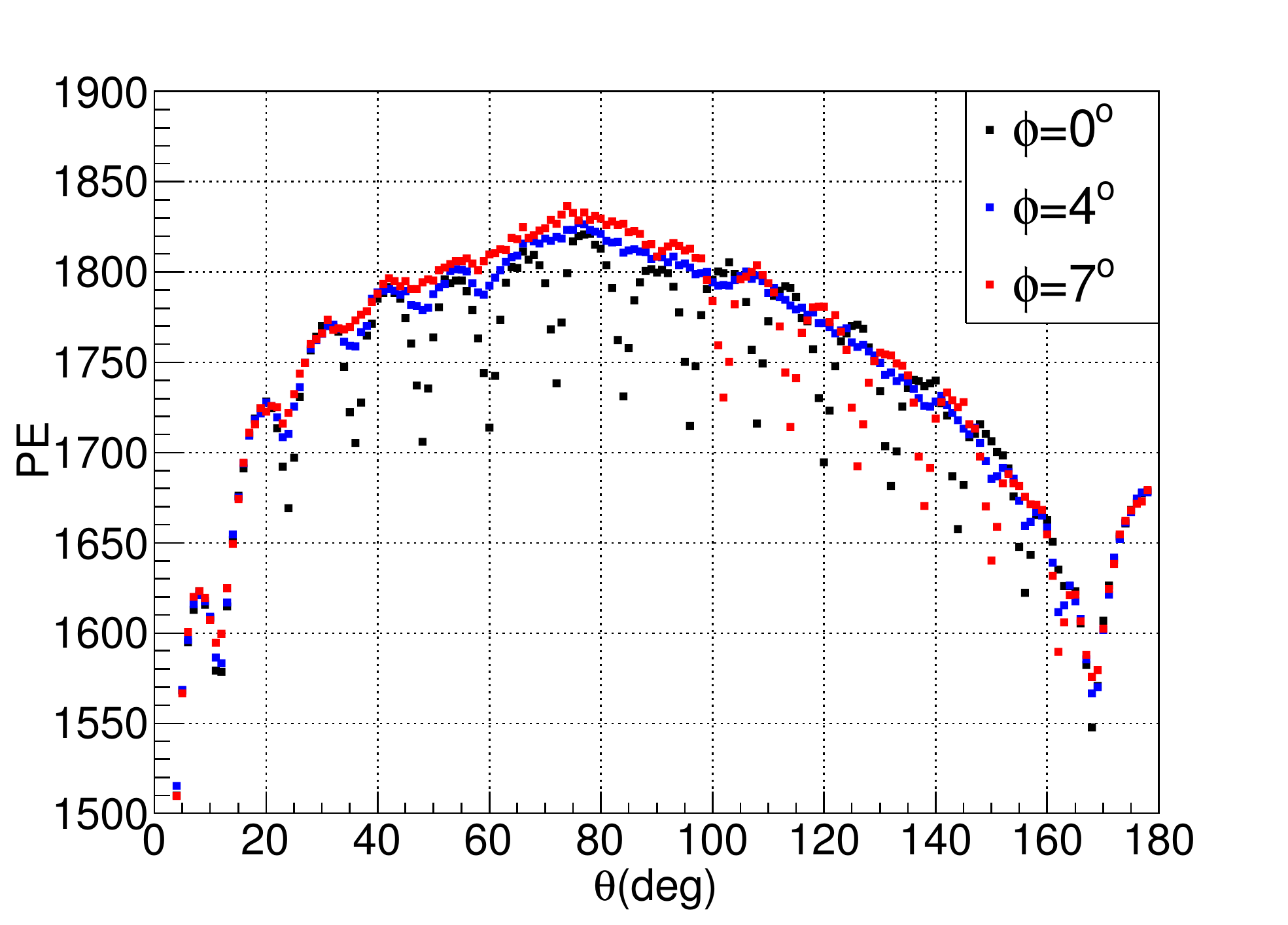}} 
    \subfigure[]{    
          \label{fig_b}     
          \includegraphics[width=0.80\textwidth]{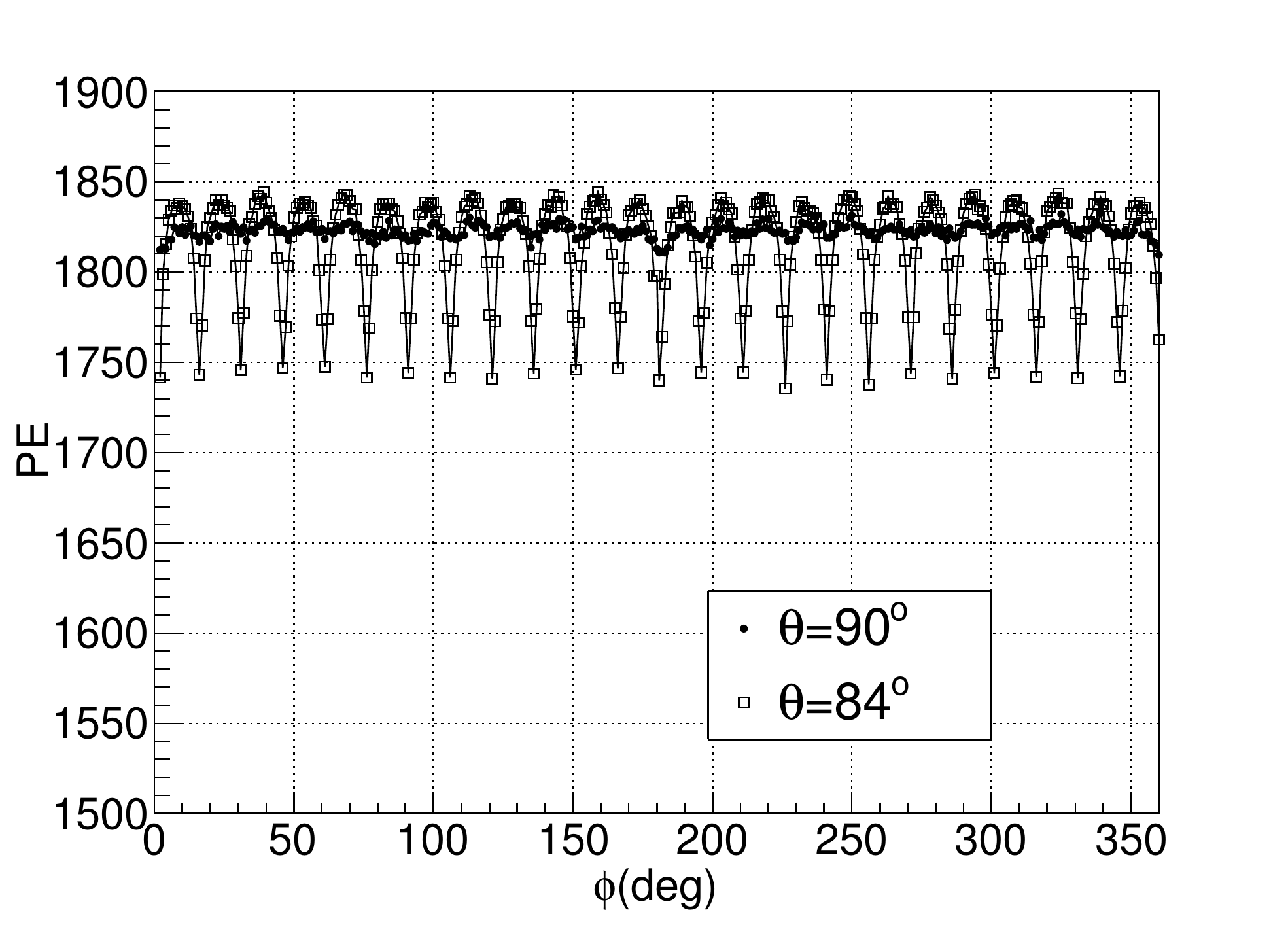}} 
	\caption{Longitude scan (a) and latitude scan (b) results.}
	\label{fig:two scan}
\end{figure*}

As mentioned above, for the region close to the CD boundary, many light blocking
structures exist, such as the SF. To settle on the location
of the GTCS, several representative scans with steps of 1\degree were
carried out along the longitude line of $\phi$=0\degree, $\phi$=4\degree,
$\phi$=7\degree and the latitude line of $\theta$=84\degree, 90\degree,
respectively. The results are shown in Fig.~\ref{fig:two scan}.
Three structures affect the response significantly. The SS chimney has
significant shadowing effects, which causes the steep slope on the
left side of longitude scan response curve in Fig.~\ref{fig:two scan} (a). The SF causes a strong local light blocking effect in the area
close to the SF (strong-SF area), as shown in Fig.~\ref{fig:two scan}
(a) with $\phi$=0\degree and 7\degree. Even at locations further away, some
residual structure can be observed (weak-SF area). This is obviously
demonstrated in the latitude response curves in Fig.~\ref{fig:two scan}
(b) with $\theta$=84\degree and 90\degree, as well as in Fig.~\ref{fig:two scan}
(a) with $\phi$=4\degree. In addition, the photon reflection on the guide tube
also affect the detector's response.

From the mechanical point of view, moving a source along the longitude
is also easier in comparison to along the latitude, e.g. around the
equator, due to a smaller friction between the source assembly and the
tube. More importantly, since the SF placement exhibits some symmetry
in $\phi$ (so is the chimney), the GTCS should be installed along the
longitude, as long as the calibration along this line would reflect
the $\phi$-averaged response. For this reason, this line should not
be too close to the SF, otherwise local light blocking effects would
be overly emphasized and bias the overall response. The goal of
the following study is to find out an efficient way to obtain the
average response function at the CD boundary.

\begin{figure*}[htp]
	\centering
    \subfigure[]{    
          \label{fig_a}     
          \includegraphics[width=0.45\textwidth]{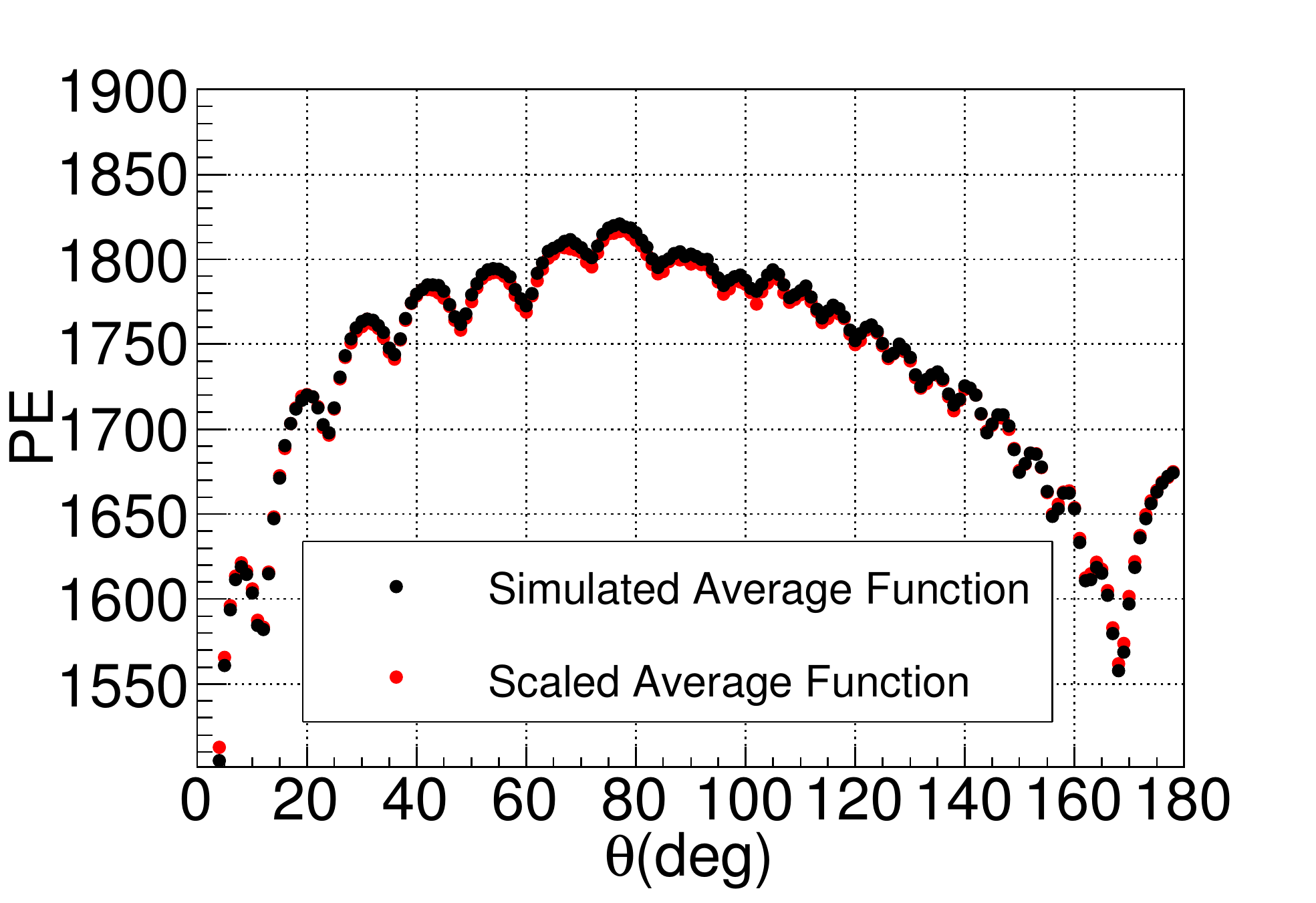}} 
    \subfigure[]{    
          \label{fig_b}     
          \includegraphics[width=0.45\textwidth]{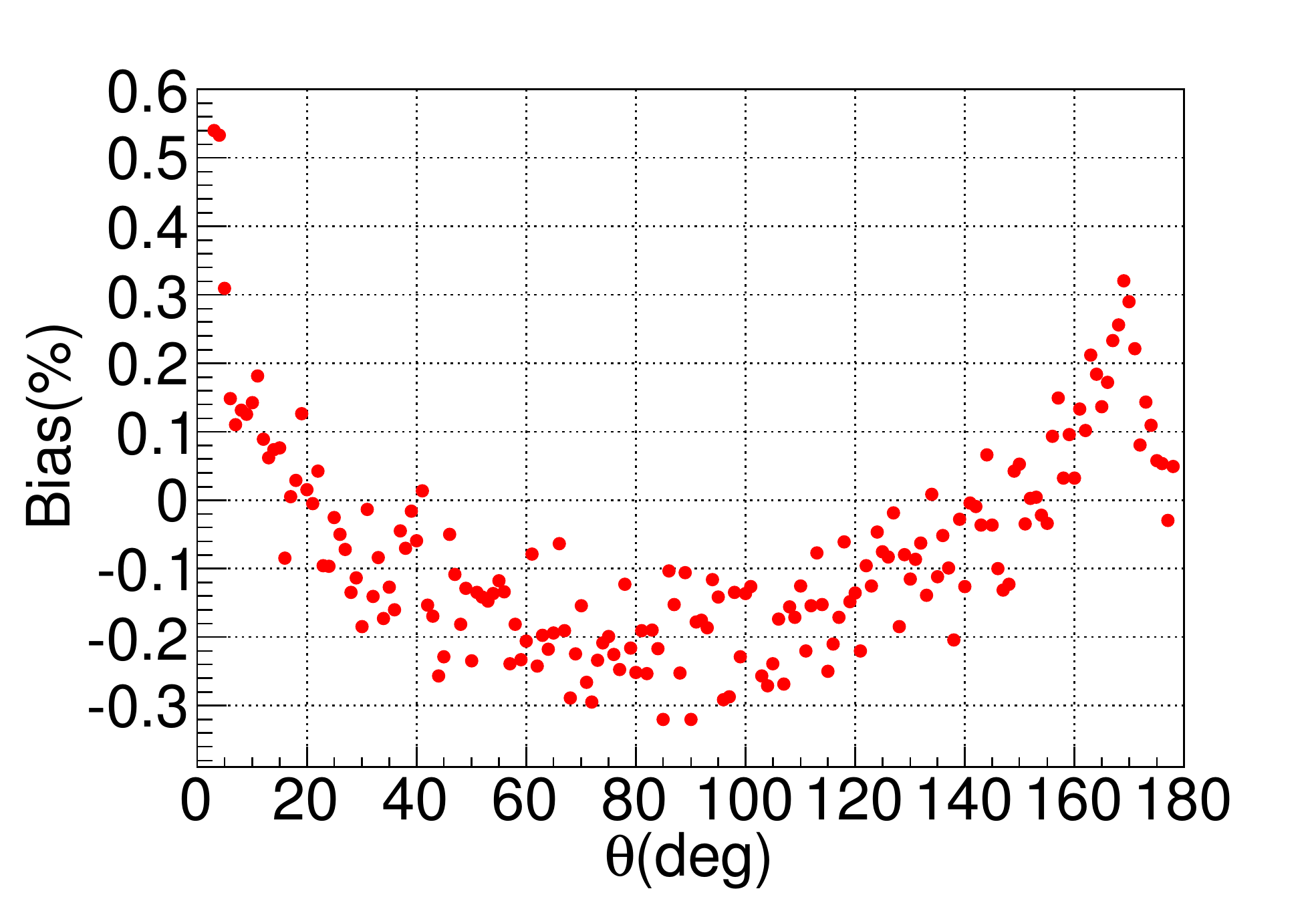}} 
	\caption{(a) The black points are the average response
          function acquired from the naked source simulation and the
          red points are the scaled response function along a
          longitude line. (b) The residual difference in fraction
          between the two in (a).}
	\label{fig:Show of the Least Square Method Correction Result}
\end{figure*}

To do this, we selected eleven longitude lines in the weak-SF area
($\theta$=30\degree, 42\degree, 54\degree, 66\degree, 78\degree,
90\degree, 102\degree, 114\degree, 126\degree, 138\degree,
150\degree), assuming the PTFE tube's reflectivity is 90\%, and
simulated the detector response. In addition, we also obtained a
$\phi$-averaged response using a perfect source from the
simulation. The comparison of these response curves is made in
Fig.~\ref{fig:Show of the Least Square Method Correction Result} (a).
The residual difference between ``calibration data'' and the
average is shown in Fig.~\ref{fig:Show of the Least Square Method
  Correction Result} (b). The overall difference is less than
0.4\%. We also applied different reflectivities in the studies and the same
conclusion is reached.

As a result, the guide tube could be installed along any longitude
line as long as the route is not blocked by a SF structure. The accurate
position will be given in next chapter with the consideration of CD
surface structures.

\subsection{Estimation of Uncertainty}
The uncertainty of the energy scale by GTCS can be classified into
four sources: a) statistic, b) FAE peak fitting, c) $\phi$-averaged
function reconstruction and d) source positioning uncertainty, in which b) and c)
have been discussed in Secs.~\ref{sec:in_or_out} and
~\ref{sec:average_function}, respectively. For a), we plan to use a
1000 Bq $^{40}$K source, which produced approximately 100/s 1.46 MeV
$\gamma$-ray. At the GTCS location, about 5\% of the events would be
FAEs.  Assuming a 5-minute nominal data taking period at each source
location, the statistical uncertainty of the FAE is estimated to be 0.1\%, 
according to Fig.~\ref{fig:The_PE_Spectrums}.

\begin{figure*}[htp]
    \centering
    \includegraphics[width=0.8\textwidth]{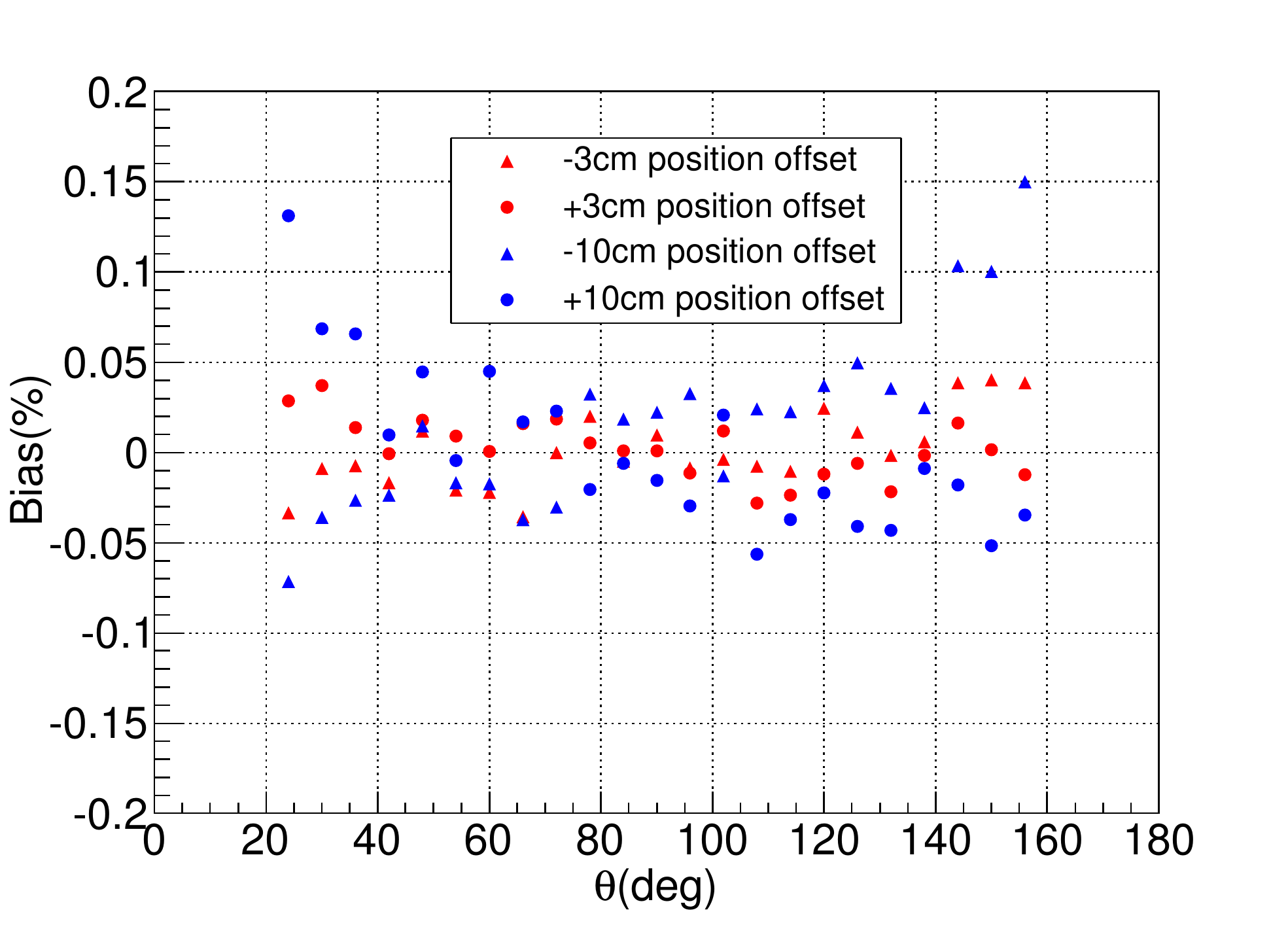}
    \caption{Response bias due to different positioning bias of the
      GTCS source.}
	\label{fig:PositionError}
\end{figure*}

To address d), we expect that the vertex reconstruction uncertainty by JUNO
photomultipliers can reach a level of 10 cm for MeV-scale positron
events. Therefore, we have required that all source locations should
be measured (independent of the reconstruction by photomultipliers) after installation. 
To study the impact of positional uncertainty on the GTCS response,
source positions were shifted intentionally between 3 cm and 10 cm 
with respect to the nominal calibration positions, and the resulting
bias to the energy response function was built using these biased positions,as shown in
Fig.~\ref{fig:PositionError}. One observes that the maximum bias is at
0.04\% level for a positional bias of 3 cm, and increases to 0.15\%
level for a bias of 10 cm. The latter bias is used as a conservative
uncertainty due to the GTCS positioning uncertainty.

As a summary, all four uncertainty sources of the GTCS is given
in Table~\ref{tab:Uncertainty}. The
overall uncertainty of the energy scale is estimated to be 0.83\%,
which meets the 1\% requirement.

\begin{table}[htbp]
\centering
\caption{\label{tab:Uncertainty} Uncertainty estimation of GTCS energy response.}
\begin{tabular}{|c|c|c|c|c|}
  \hline
  Statistical & Positional bias& FAE fitting & Average response reconstruction & Total \\ \hline
  0.1\% & 0.15\% & 0.7\% & 0.4\% & 0.83\%\\\hline
\end{tabular}
\end{table}

\section{GTCS Mechanical Design}
\label{sec:mechanical}
\begin{figure*}[htp]
	\centering
	\includegraphics[width=0.9\textwidth]{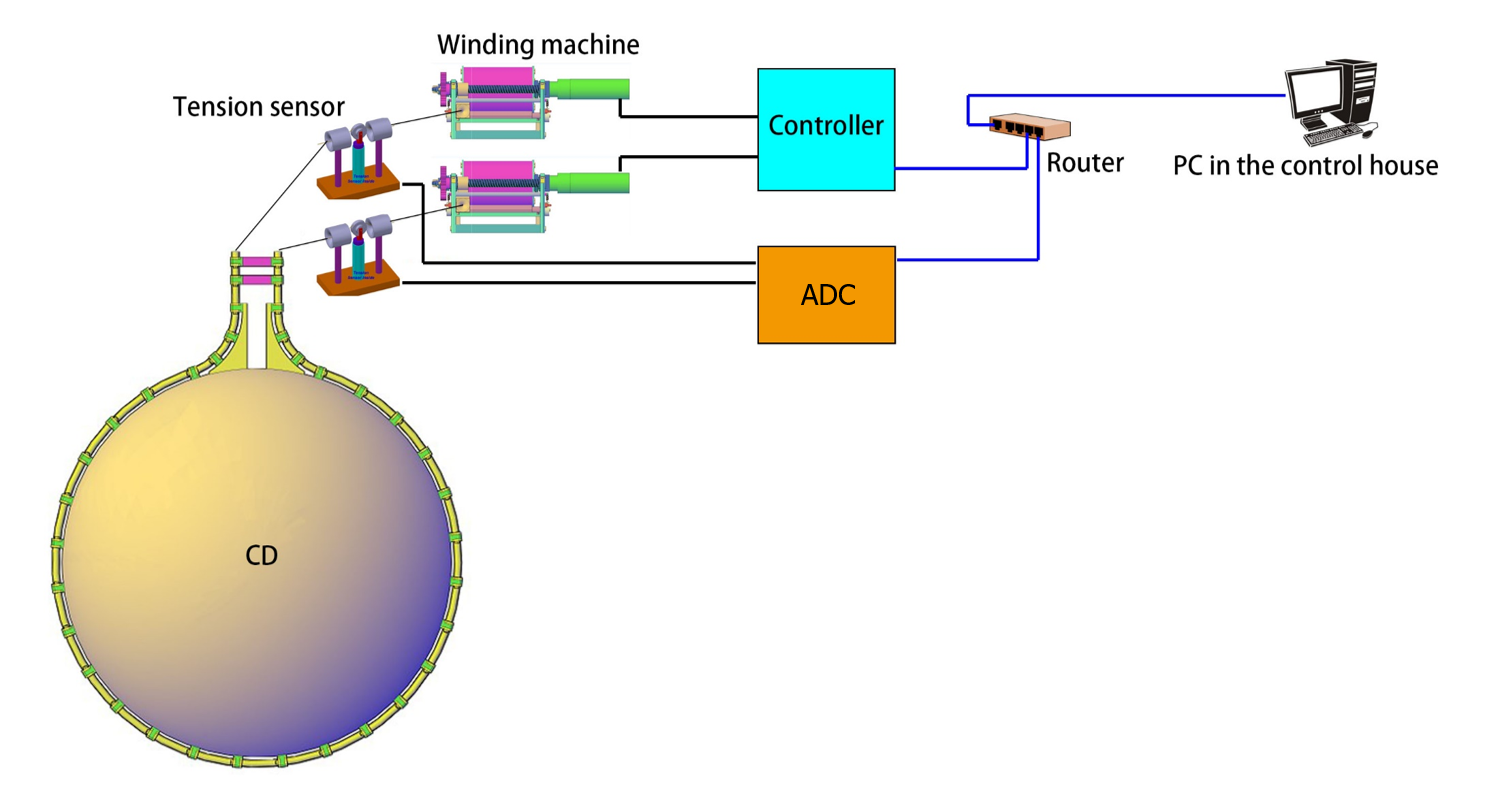}
	\caption{A general diagram of GTCS.}
	\label{fig:GTCS_general}
\end{figure*}

Based on the study above, we now discuss the mechanical design of the
GTCS.  Our primary purpose is to deliver the radiation source to the
designated positions and later to retract it safely. The system is
divided into three parts: (1) the source and cable assembly; (2) guide
tube and its fixing anchors on the CD; (3) winding machine and its
control system.  A general diagram of GTCS is shown in
Fig.~\ref{fig:GTCS_general}.

\subsection{Source and Cable Assembly}
\begin{figure*}[htp]
	\centering
	\includegraphics[width=0.4\textwidth]{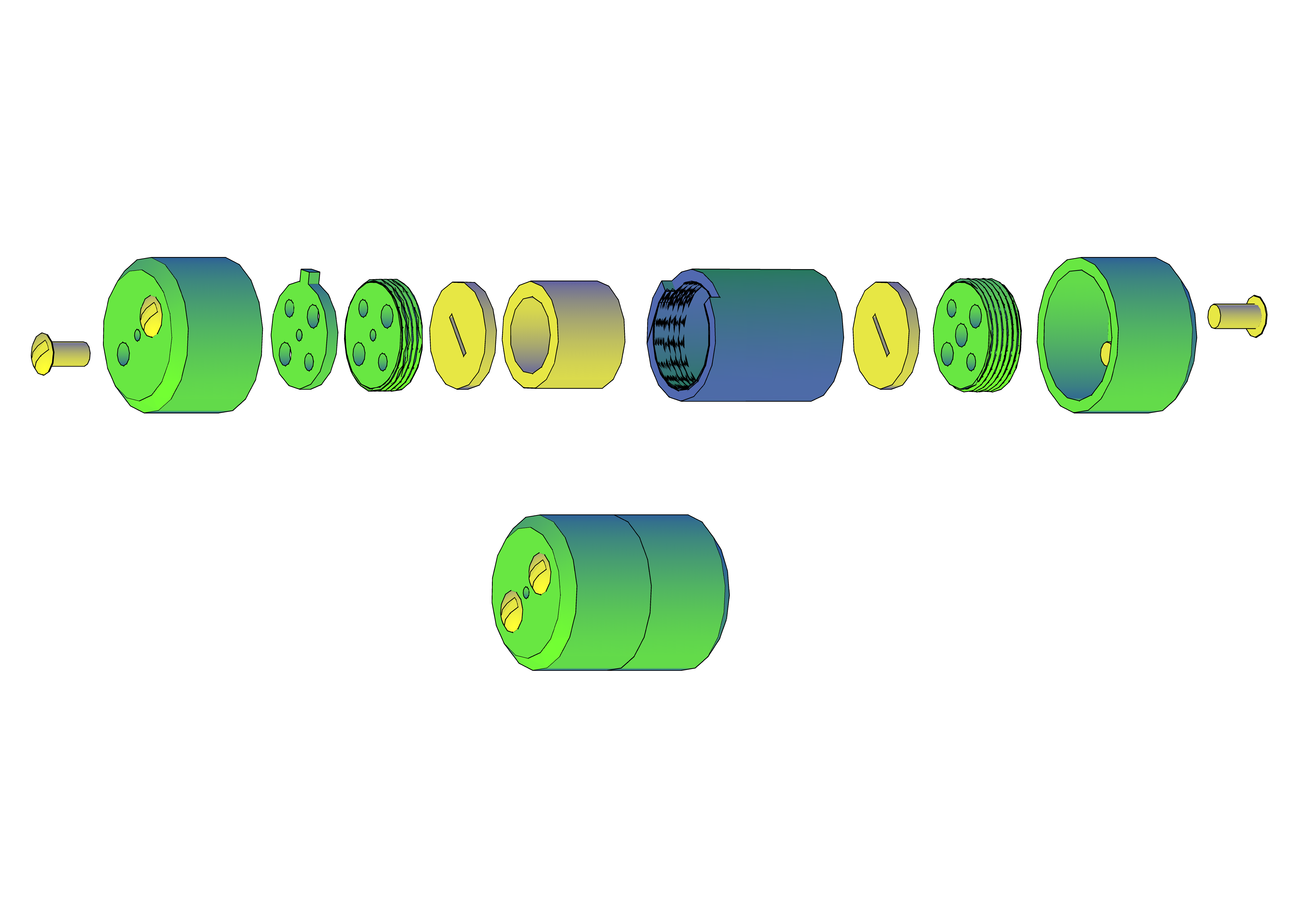}
	\caption{The design of the source assembly.}
	\label{fig:The Scheme of source carrier}
\end{figure*}

The design of the radioactive source assembly is illustrated in
Fig.~\ref{fig:The Scheme of source carrier}.  The calibration source
will be sealed in a SS cylinder ($\phi6\times 6$). The enclosure of
the  cylinder is made out of a PTFE cylinder with an overall dimension of
$\phi13\times20.5$, in order to minimize friction and turning radius. Two 130-m-long cables are attached on both ends of
it to move the source assembly in two directions. This design also
offers extra insurance if either the source is stuck in the tube or
one cable is broken. The cable is a SS wire with a diameter of 0.7~mm
coated by a 0.15-mm-thick FEP. A tensile test indicated its minimum
breaking force is greater than 800~N.

\subsection{The Tube and its Fixing Anchors}
The main body of GTCS is a 130-m-long PTFE tube with inner diameter of
16~mm and a wall thickness of 1.5 mm. PTFE is chosen to minimize friction. The tube should be
deployed along a longitude of acrylic sphere and then go along the
JUNO chimney to get out of the water shield.

\begin{figure*}[htp]
	\centering
	\includegraphics[width=0.5\textwidth]{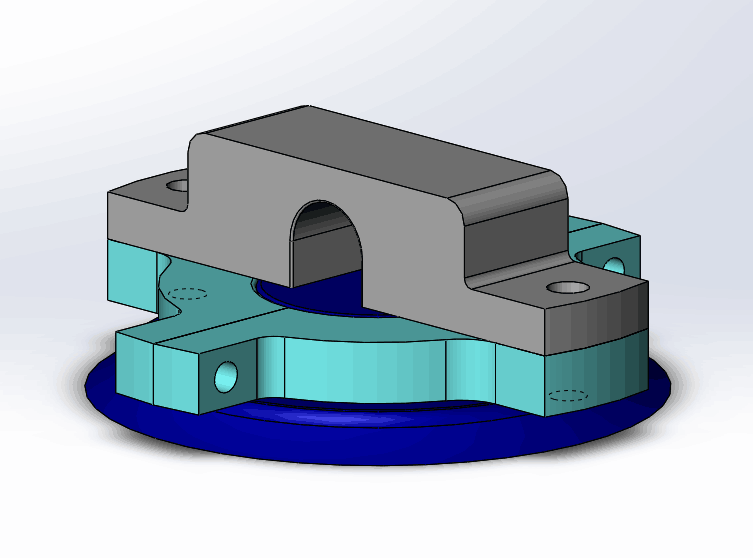}
	\caption{The design of the GT anchor outside the CD.}
	\label{fig:GTanchor}
\end{figure*}

The fixture of the tube on CD surface is also called the GT anchor. As shown
in Fig.~\ref{fig:GTanchor}, it is composed of three parts. The dark
blue part is an acrylic base and it will be bonded onto the CD surface
via polymerization bonding method. The shape is carefully designed to
minimize residual stress during bonding. Both the cyan and grey parts
are adjustable nylon holder for the tube. The tube will go through the hole in
grey holder which will be fixed on the cyan holder using bolts. 

\begin{figure*}[htp]
	\centering
	\includegraphics[width=0.6\textwidth]{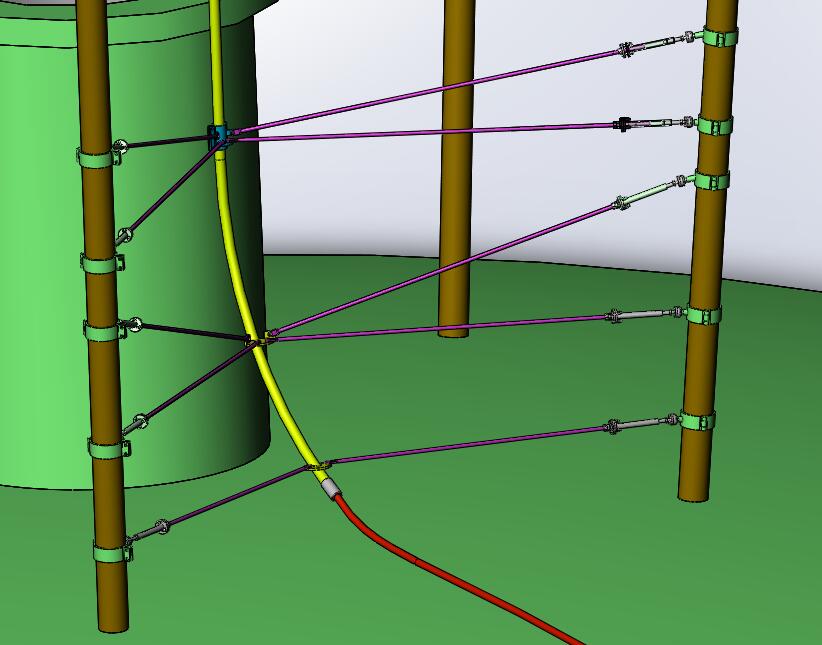}
	\caption{Details of the turning track for the GT.}
	\label{fig:The Scheme of Turning Track}
\end{figure*}

On the CD, there are in total 58 anchors, about one in every two meters. 
Near the CD chimney, a turning track with a radius of 2 m made of SS
tube is fixed on the SS connection bars with  rods, as shown in 
Fig.~\ref{fig:The Scheme of Turning Track}.

\begin{figure*}[htp]
	\centering
    \subfigure[]{    
          \label{fig_a}     
          \includegraphics[width=0.3\textwidth]{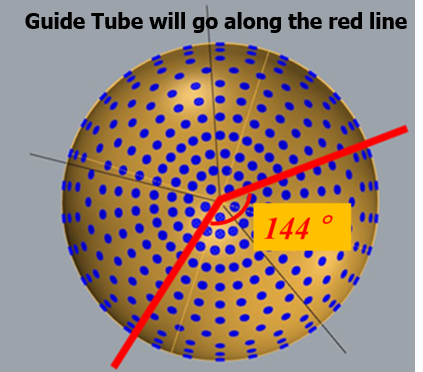}} 
    \subfigure[]{    
          \label{fig_b}     
          \includegraphics[width=0.4\textwidth]{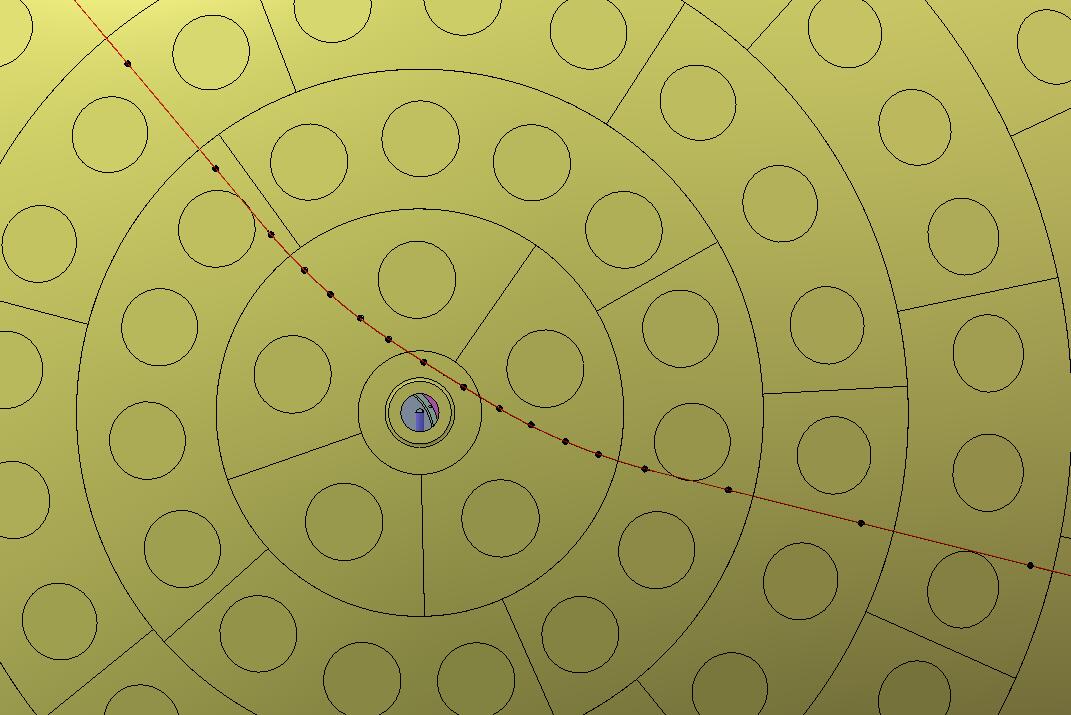}} 
	\caption{(a) The GT and its anchors on the CD surface; (b) the
          routing of the GT close to the bottom of the CD. The SF
          placement has a 5-fold symmetry in $\phi$ and the $\phi$
          angle between two half circles of GT is 144\degree.}
	\label{fig:CD_nodes}
\end{figure*}

As discussed in Section ~\ref{sec:average_function}, the guide tube could
be installed along any weak-SF longitude line to represent a
$\phi$-averaged boundary response. To avoid conflict with the SF
structure, the whole GT circle is divided into two half circles: one
is on the longitude of 2.61\degree and the other one is on the
longitude of 146.61\degree, as illustrated in
Fig.~\ref{fig:CD_nodes}. At the bottom of CD, the PTFE tube detours
smoothly from the bottom flange of the CD.

The GT will be immersed in the ultrapure water shield with a
maximum height of 40~m.  To avoid deformation due to water pressure, the
guide tube is designed to be filled with water, which will be
recirculated to ensure the cleanliness.

\subsection{Winding Machine and its Control System}

\begin{figure*}[htp]
	\centering
    \subfigure[]{    
          \label{fig_a}     
          \includegraphics[height=4cm]{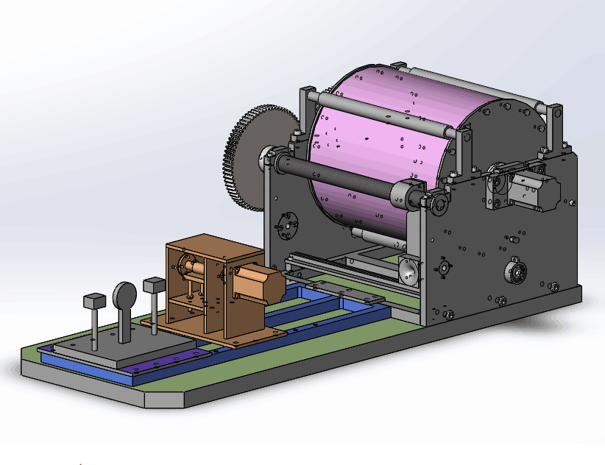}} 
    \subfigure[]{    
          \label{fig_b}     
          \includegraphics[height=4cm]{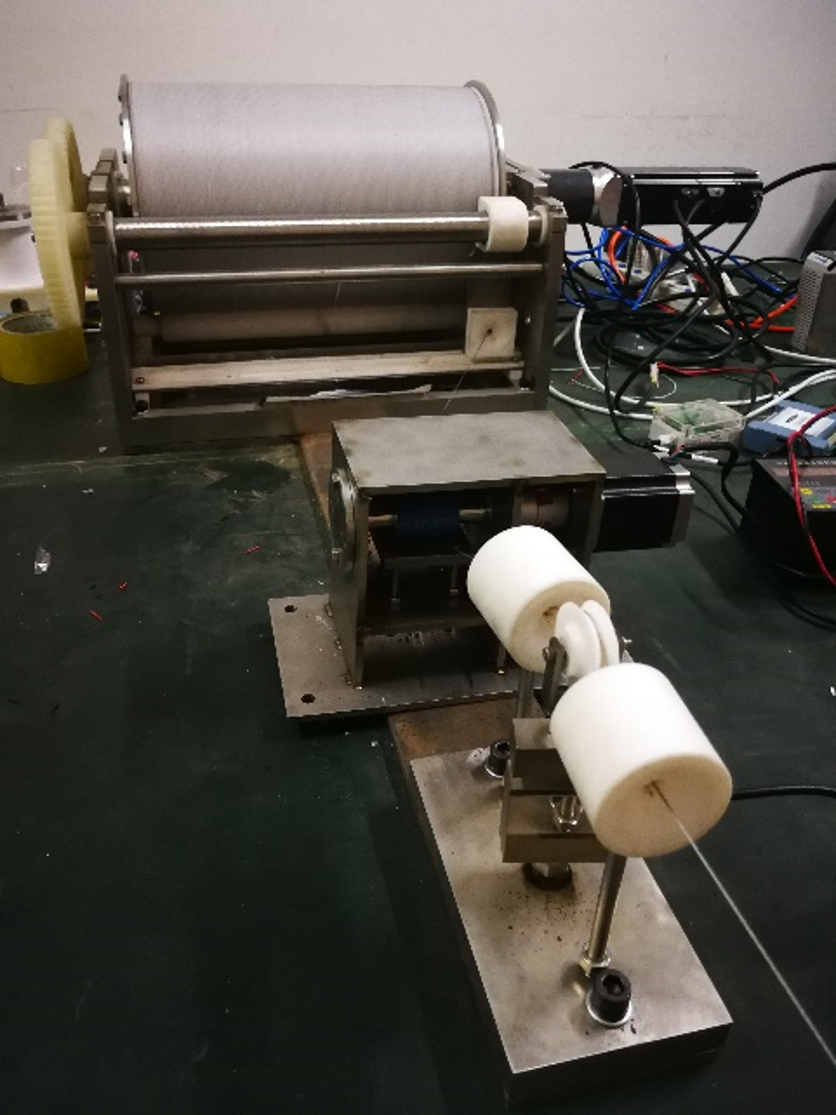}} 
    \subfigure[]{    
          \label{fig_c}     
          \includegraphics[height=4cm]{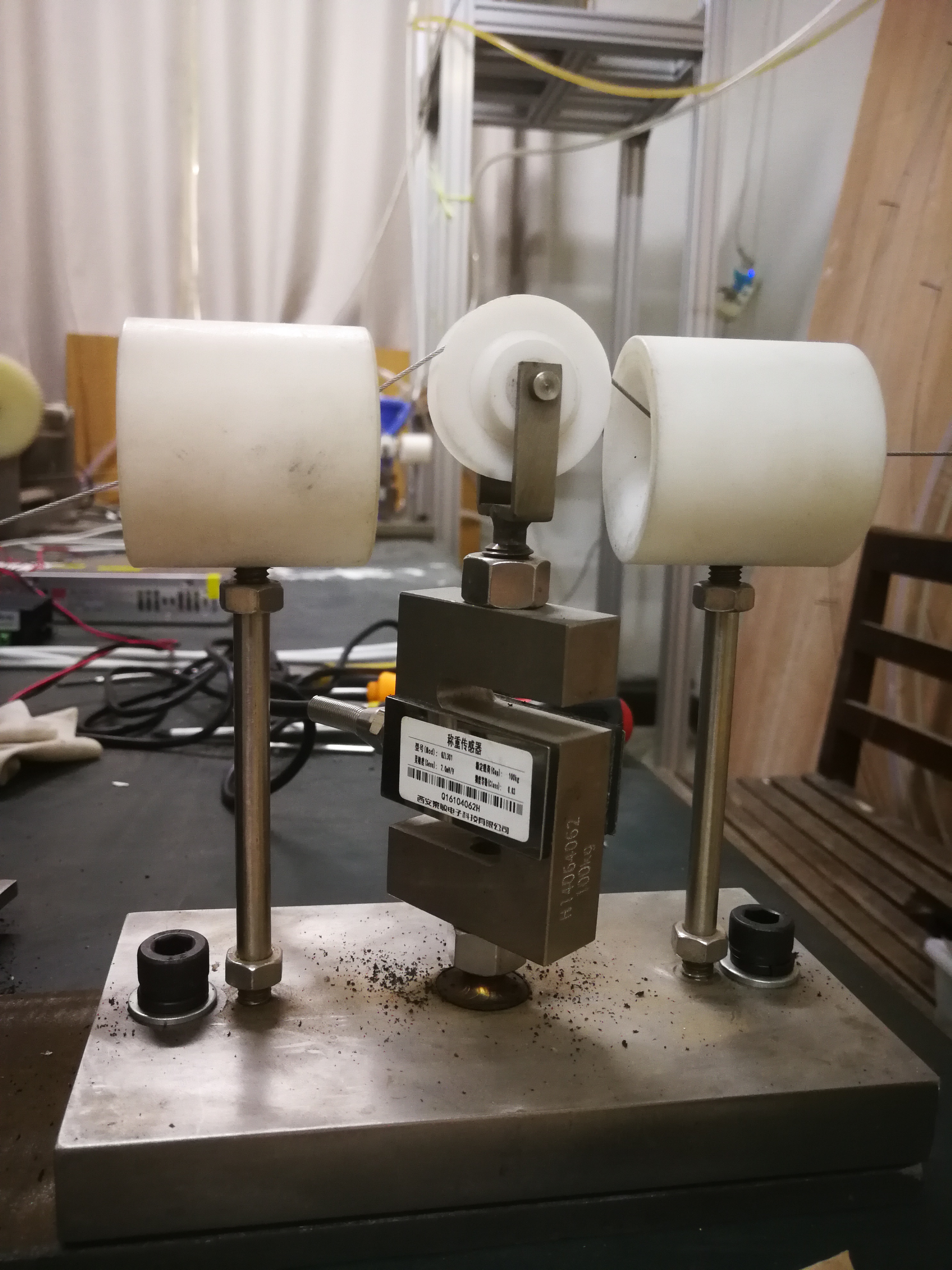}} 
	\caption{Design drawings of the winding machine (a) and (b),
          and the tension pulley (c).}
	\label{fig:tension_sensor}
\end{figure*}

The winding machine is the key part of GTCS. It will control the
source assembly position by dragging the deployment cables, as shown
in Figs.~\ref{fig:tension_sensor} (a) and (b). Its main body is a PTFE
wire spool with helical grooves on which the cable is wound in
order. A servo motor, controlled by a LabVIEW software, drives the
spool automatically.  The tension in the cable is monitored constantly
to avoid the source getting caught unexpectedly in the GT.  To achieve
this, a load cell is mounted below a pulley through which the cable
goes through with an angle of 120\degree \ before the spool
(please see Fig.~\ref{fig:tension_sensor} (c)). 

The winding spool obtains the source location via the length of
cables. Although guide tube is fixed onto the  anchors approximately every two meters, the
distortion of the guide tube could still introduce a positioning
uncertainty. To better control this, metal proximity sensors will be
installed close to a few selected CD anchors. A signal will be
triggered when the source assembly goes by a sensor, with a sub-cm
precision based on a prototype test. With these sensors, the source
positioning uncertainty for the entire GT loop is expected to reach
the 3-cm requirement.

\begin{figure*}[htp]
	\centering
	\subfigure[]{    
		\label{fig_a}     
		\includegraphics[height=6cm]{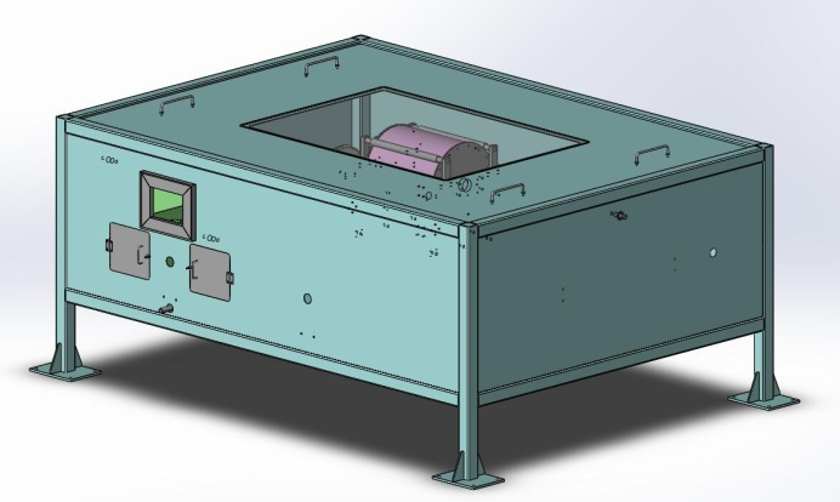}} 
	\subfigure[]{    
		\label{fig_b}     
		\includegraphics[height=6cm]{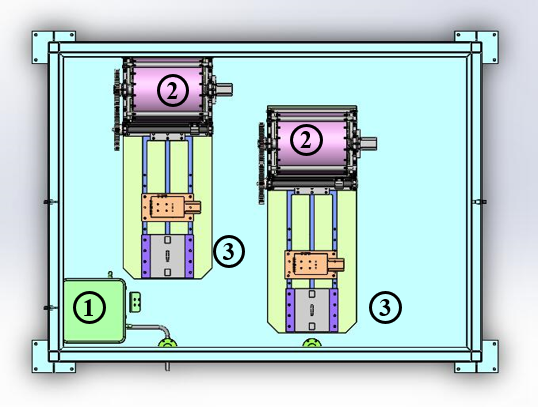}} 
	\caption{(a) External appearance of the GTCS box. (b) Internal
		layout of GTCS box: (1) tank of the water recirculating
		system (2)winding machines (3) cable tension pulleys.).  }
	\label{fig:GTCShouse}
\end{figure*}

The winding machines and related auxiliaries will be enclosed in a
GTCS box, as shown in Fig.~\ref{fig:GTCShouse}, which is located on a
bridge above the water and close to the chimney. To avoid light
leakage into the water and support the guide tude vertically, a SS tube enclosing the guide tube above the water is connected to turning track and goes to the GTCS box with a sealed penetration. 

Our mechanical design  was verified under realistic conditions.
While details of this test will be reported elsewhere, we comment on
the general performance here.  We constructed a full-size circular
track, horizontally laid on the ground with a diameter of 35.4~m. The
guide tube was attached to the track. The motion of the source
assembly was automatically controlled and tested, showing the required
precision could be satisfied. The maximum friction between source assembly and tube was measured to be less
than 20 N, which is very low compared to the break strength of 800~N. 

\section{Summary}
In summary, we have designed a guide tube calibration system for the
JUNO experiment. Based on a realistic simulation, such a system is
capable of calibrating the energy scale at the CD boundary to a
sub-percent level. The mechanical design of the GTCS has beenm developed and
verified to full scale.

\acknowledgments

We thank Tao Zhang from SJTU, Xiaoyan Ma and Xiaohui Qian from IHEP for engineering support
in the mechanical design. This research is supported by the "Strategic
Priority Research Program" of the Chinese Academy of Sciences (Grant
No. XDA10000000).


\end{document}